\documentclass{aa}  

\usepackage{graphicx}
\usepackage{hyperref}
\usepackage{txfonts}
\usepackage{booktabs}
\usepackage{amsmath}
\usepackage{amssymb}
\usepackage{textcomp}
\usepackage{xcolor}

\begin{document} 
\title{Kinematics of the lens host S0 galaxy NGC 1553: role of secular processes}

   \author{Saili Keshri
          \inst{1}\fnmsep\inst{2} 
          \and
          Sudhanshu Barway
          \inst{1}
           \and
          Francoise Combes
          \inst{3}
          }
   \institute{Indian Institute of Astrophysics, Kormangala II Block, Bengaluru, India, 560034\\
            \email{saili.keshri@iiap.res.in}
        \and
            Department of Physics, Pondicherry University, R.V.Nagar, Kalapet, Puducherry, India, 605014
        \and
            Observatoire de Paris, LUX, College de France, CNRS, PSL University, Sorbonne University, F-75014 Paris, France\\
        }
\abstract {We present an investigation of the central structure of the S0 galaxy NGC1~553, to understand its origin and the underlying dynamical processes that shape it. The high-resolution integral field spectroscopic data from the Multi-Unit Spectroscopic Explorer (MUSE) reveal a well-ordered rotation pattern, consisting of a rapidly rotating nuclear disc that is somewhat decoupled from the main disc, together with an inner lens; we collectively refer to these structures as the "disc-lens". The central peak in the velocity dispersion indicates the presence of a classical bulge. The nuclear disc is dynamically colder than the surrounding disc, while the lens is dynamically hotter. The higher-order Gauss–Hermite moments, $h_{3}$ and $h_{4}$, further characterise the stellar kinematics. An anti-correlation between the line-of-sight velocity and skewness ($h_{3}$) is consistent with regular rotation. In contrast, the ring-like enhancement in kurtosis ($h_{4}$) confirms the presence of the nuclear disc component. Unsharp masking of \textit{HST} images \citep{Erwin15} reveals a nuclear bar and faint spiral structures within the central $\sim$10~arcsec, supporting the role of secular evolution. The mass-weighted stellar age map shows an old stellar population in the central regions, with high metallicity that suggests the in-situ formation of the disc–lens from disc material. We discuss possible formation scenarios for the disc–lens, including both minor mergers and secular processes, and examine the influence of the Dorado group environment on NGC~1553. Our findings suggest that the disc–lens in NGC~1553 formed during the early stages of the galaxy's evolution. However, its subsequent development has been shaped by internal and external processes. These results provide new insights into the origin and evolution of kinematically distinct substructures in S0 galaxies.}

\maketitle

\section{Introduction}\label{sec:Sec_1}
S0 galaxies occupy a crucial transitional position in the Hubble sequence between elliptical and spiral galaxies, exhibiting a range of structural components such as bulges, nuclear discs/rings, bars, and lenses \citep{Kormendy04, Laurikainen13}. These features offer insights into the evolutionary processes that shape galaxies, including both external interactions and internal secular evolution. The nuclear discs are generally built through bar-driven processes that shape the main galaxy discs \citep{Falcon04, Falcon06}. They host typically inner bars, nuclear spiral arms, and a nuclear ring that forms at the outer rim of the nuclear disc \citep{Cole2014}. Numerical simulations have also shown that nuclear discs can occur in unbarred galaxies, where their formation originates from the accretion of external gas or mergers, in processes unrelated to bars \citep{Mayer10, Chapon13}. However, nuclear discs in unbarred galaxies are typically less extended than those formed through bar-driven processes. Another key structure observed in many S0 galaxies is the presence of a lens, a component with a nearly uniform surface brightness and a sharp outer edge, whose formation mechanisms are yet to be understood. The interplay of dynamical processes, such as bar evolution, minor mergers, and environmental influences, contributes to the development of lenses and nuclear substructures in these galaxies \citep{Buta10, Eliche-Moral18}. 

One widely accepted scenario suggests that lenses are the remnants of dissolved or weakened bars \citep{Kormendy79, Laurikainen11}. Over time, bars can evolve due to dynamical friction, gas inflow, and resonant interactions, leading to a redistribution of stellar orbits that results in a smooth, lens-like structure \citep{Buta10}. When the remnant of a dissolved bar, the outer boundary of a lens corresponds to the corotation radius, often exhibiting an enhancement similar to that of a ring \citep{Combes96}. Alternatively, numerical simulations have shown that lenses can also arise from minor mergers, where the accretion of small satellite galaxies leads to the redistribution of stars and the formation of a lens-like component \citet{Buta10, Eliche-Moral18}.

In barred galaxies, the presence of a lens is often linked to the evolution of the bar itself. Bars can undergo secular transformations through angular momentum exchange with the surrounding disc and dark matter halo, forming an oval lens structure that retains the original bar's orientation \citep{Combes96, Athanassoula16}. This process is more pronounced in lenticular galaxies, where the lack of a significant gas reservoir prevents continuous star formation, allowing structural evolution to proceed more efficiently. Additionally, some lenses may form due to resonances associated with bar dynamics, where stellar orbits are trapped in specific configurations that create a nearly uniform surface brightness distribution \citep{Kruk18}.
\begin{figure*}
    \hspace{0.3cm}
    \centering
    \includegraphics[width=16cm,height=8cm ]{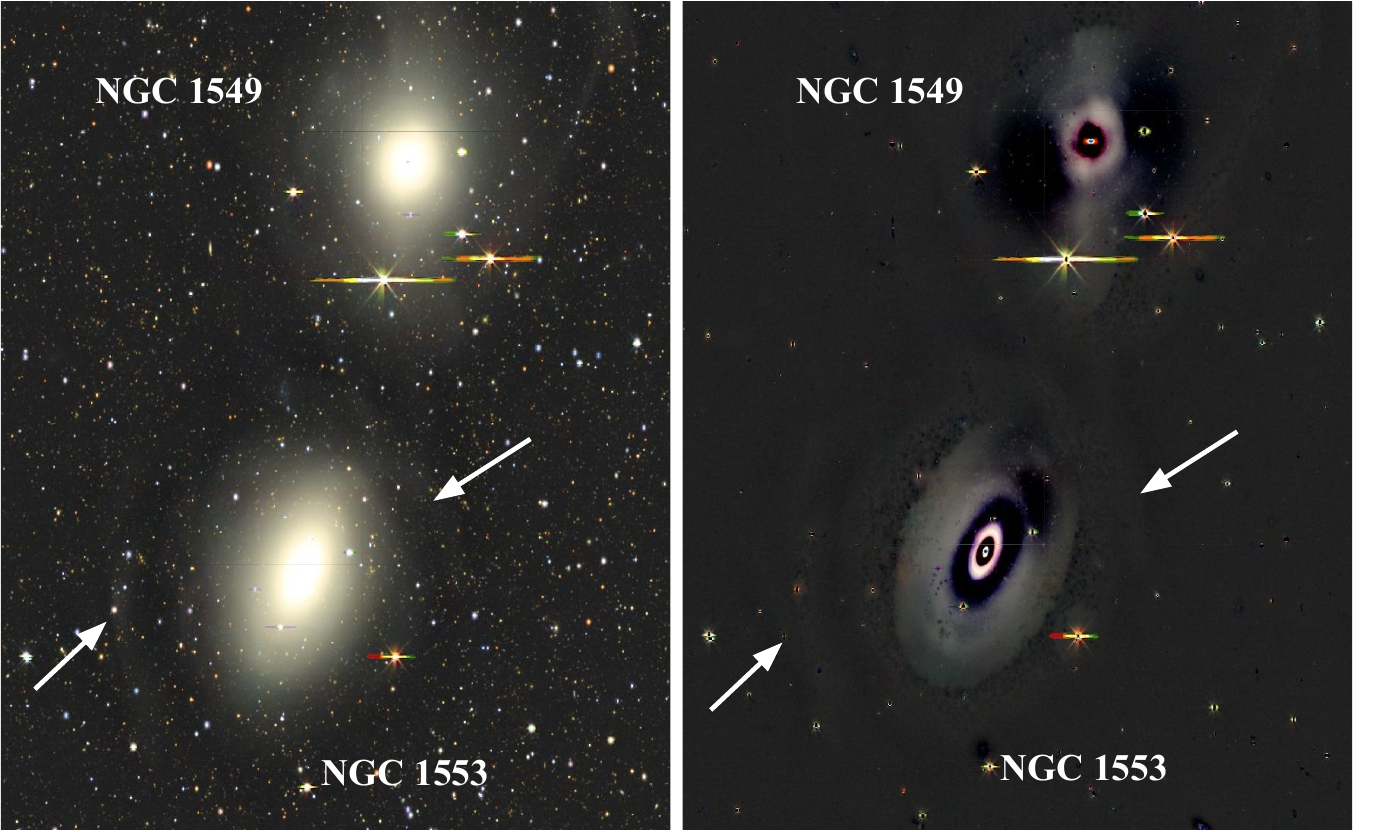}
    \caption{DECaLS color composite image (left) and residual image (right) of NGC 1553 and nearby galaxy NGC 1549. The white arrow in both images represents the faint shell feature in NGC 1553. North is oriented upward, and east is to the left. The size of the bow is 22 arcmin in each panel.}
    \label{fig:decals}
\end{figure*}

S0 galaxies provide a unique environment for studying lens formation due to their intermediate nature between spirals and ellipticals. Observational studies have revealed that lenses are significantly more common in S0 galaxies than in spirals. It supports the hypothesis that their formation is closely linked to the transformation processes that shape these galaxies \citep{Laurikainen13, Gao18}. In many cases, the presence of a lens is accompanied by a faded bar, indicating a secular evolution pathway where the original bar gradually dissolves into a dynamically hot component \citep{Shen04}.

Another important aspect of lens formation in S0 galaxies is the role of environmental interactions. Since S0 galaxies are frequently found in galaxy groups and clusters, they are more susceptible to mechanisms such as galaxy harassment, tidal stripping, and minor mergers, which can contribute to the formation and evolution of lens structures \citep{Bekki11, Wilman09}. Shells, rings, and tidal features observed in many S0s, including NGC 1553 studied here, suggest a history of interactions that likely played a role in shaping their central structures\citep{Malin83, Gourab23}.

In the present work, we use high-resolution integral field spectroscopic (IFU) data from Multi Unit Spectroscopic Explorer (MUSE) to conduct a comprehensive analysis of the central structure of NGC 1553.
The paper is organised as follows. In Sec~\ref{sec:properties} we describe the photometric properties of NGC 1553. Sec ~\ref{sec:Data} provides a brief overview of the data. The analysis is presented in Sec~\ref{sec:muse analysis}. We present our results in ~\ref{sec:results}, followed by a detailed discussion and conclusion in Sec ~\ref{sec:discussion}. Throughout this work, we adopt a flat $\Lambda$CDM cosmology with $\Omega_M$ = $0.3$, $\Omega_{\lambda}$ = $0.7$ and Hubble constant (H$_0$) = $70km$ $s^{-1}$ $Mpc^{-1}$.

\section{Photometric properties of NGC 1553}\label{sec:properties}
NGC 1553 is classified as an unbarred lenticular (S0) galaxy, exhibiting a prominent lens or inner ring structure \citep{Sandage79, Kormendy84}. In the Third Reference Catalog of Bright Galaxies (RC3), this feature is designated as an inner ring \citep{deVaucouleurs91}. \citet{Laurikainen06} identified a nuclear disc with spiral arms in the central region of the galaxy. They determined a bulge-to-total luminosity ratio ($B/T$) = 0.21 and a Sersic index = 1.9, indicating that the bulge exhibits characteristics consistent with a pseudo-bulge. The pseudo-bulge character is reinforced by its high V/$\sigma$ value \citep{Kormendy04}. Using data from the S$^{4}$G survey, \citet{Buta15} classified the galaxy as an unbarred S0 system, exhibiting a ring–lens, a nuclear ring or lens, and a nuclear bar. The inner morphology is particularly complex, and a detailed kinematic analysis of the central region with high-resolution data is required to better constrain and characterise these components. The galaxy also hosts a prominent hot lens \citep{Kormendy84}, which appears as a shelf in the surface-brightness profile between 20 and 36 arcsec. However, the inner part of the lens begins to dominate the surface-brightness profile at $\sim$12–15 arcsec \citep{Kormendy84}.

Further evidence for a disc-like bulge was provided by \citet{Erwin15}, who identified a pseudo-bulge within the central region (r < 16 arcsec) of NGC 1553. High-resolution HST imaging further revealed the presence of a nuclear stellar bar surrounded by spiral arms, reinforcing the classification of a disc-like bulge. Additionally, \citet{Erwin15} reported that the isophotes within the nuclear bar region appear rounder, accompanied by an excess in the central surface brightness profile. A 2D decomposition of the inner r < 30 arcsec region, employing a Sersic + exponential profile, demonstrated that the inner Sersic component dominates at r = 1.48 arcsec, with a Sersic index of 1.66, which corresponds to the round central isophotes. Combining these photometric properties with kinematic studies \citep{Kormendy84, Longo94}, \citet{Erwin15} classified the inner photometric bulge as a classical bulge. Consequently, NGC 1553 is an example of a galaxy hosting a composite bulge structure comprising both classical and pseudo-bulge components.

NGC 1533 is also classified as a shell galaxy. The shell was revealed by applying a photographic amplification technique\citep{Malin78}, which enhances the faint features \citep{Malin83}. The past merger event had produced faint but distinct shells and a curious central spiral structure. The galaxy also appears to be at an early stage in an encounter with neighbouring elliptical NGC 1549. The projected separation between the two galaxies is 11.8 arcmin ($\sim$ 63.5 kpc) \citep{Bridges90}.
NGC 1553 is part of the Dorado group, which contains five galaxies as suggested by \citet{Garcia93}. Later HST images reveal an inner torus-like dust lane in the central 3~arcsec, which was detected in the infrared in all four IRAS bands. It is also a weak radio source with a flux density of 10 mJy at 843 MHz. The $H{\alpha}$ narrow band image of NGC 1553 shows a strong nuclear peak and spiral-like structure at 8 arcsec from the nucleus \citep{Trinch97}. A recent study by \citep{Ricci23} on the nuclei of early-type galaxies detected the presence of $H\alpha$, $H\beta$ and a broad-line region (BLR) in the nucleus of NGC 1553, indicating type 1 AGN/LINER-type emission.

The X-ray Chandra observation by \cite{Blanton01} reveals that 70\% of the emission in the 0.3-10 keV band is diffuse, while the remaining arises from discrete sources (low-mass X-ray binaries). The diffused emission is predominantly soft, mainly from the thermal emission from hot gas. A very bright source coincides with the optical nucleus; its spectrum and luminosity are consistent with it being an obscured active galactic nucleus (AGN). Chandra observations detect spiral features in the diffuse emission passing through the centre of the galaxy, probably due to an adiabatic or shock compression of the ambient gas, but not due to cooling. 

\section{Data} \label{sec:Data}
We used archival optical IFS data obtained with the MUSE \citep{Bacon06}) mounted on ESO's Very Large Telescope (VLT). For this study, we analysed the MUSE wide field adaptive optics mode (WFM-AO-N) data cube of the galaxy NGC 1553 to investigate its stellar kinematics \& population. Observations were conducted over the nights of 2019 December 22–23 under programme ID 0104.B-0404. The science-ready data cube was retrieved from the ESO Science Archive Portal\footnote{\url{https://archive.eso.org/scienceportal/home}}.
The observations span a spectral range of 470–935 nm, with a median spectral resolution (R) of 3027, enabling deblended identification of many important optical emission lines. The field of view of the data cube covers an area of 1.09 arcmin $\times$ 1.07 arcmin, with a total integration time of 1760 seconds. The effective seeing during the observations was approximately 1.059 arcsec. 
To analyse the MUSE data cube, we followed a methodology similar to that described by \citep{Saili25}.

In the left panel of Figure~\ref{fig:decals}, we show the color composite image of NGC 1553, while the right panel displays the corresponding residual image, both images are taken from the Dark Energy Camera Legacy Survey (DECaLS, \citep{Dey19}). NGC 1553 is interacting with NGC 1549, located 11.8 arcminutes ($\sim$ 63.5 kpc) away. A faint shell in NGC 1553, marked by a white arrow, is visible in both images. The residual image also reveals an inner ring. At the adopted distance of NGC 1553 (18.5 Mpc, based on \citet{Blakeslee01}), an angular scale of 1 arcsecond corresponds to a physical scale of 89.7 parsecs.

\begin{table}[h]
	\centering
        \caption{Properties of NGC 1553. The morphology is taken from \citet{Buta15} and the distance is adopted from \citet{Blakeslee01}. Right Ascension and Declination of the galaxy are taken from NASA/IPAC Extragalactic Database2. Galaxy inclination and disc position angle are adopted from \citet{Erwin15}.}
        \begin{tabular}{lr}
        \hline
        Parameter & Value \\
        \hline
        \hline
        Morphology & SA(rl,nrl,nb)0+ \\
        Distance (D) [Mpc] & 18.5 \\
        Right Ascension (J2000) [Deg] & 64.043484 \\
        Declination (J2000) [Deg] & -55.780005 \\
        Inclination angle [Deg] & 48 \\
        Position angle of disc photometric axis [Deg] & 152 \\
        \hline       
        \end{tabular}
	\label{tab:Table_1}
\end{table}



\section{MUSE data cube analysis}\label{sec:muse analysis}

We employed version 3.1.0 of the Galaxy IFU Spectroscopy Tool (GIST; \citealt{Bittner19}), a comprehensive analysis framework designed for fully reduced IFU data, to process the MUSE data cube. Stellar kinematics were extracted using the penalized pixel-fitting (pPXF) algorithm \citep{Cappellari04, Cappellari17}, which fits the stellar line-of-sight velocity distribution (LOSVD). Emission line fluxes and gaseous kinematics were derived using the Python implementation of Gas and Absorption Line Fitting (GandALF; \citealt{Sarzi06, Falc'on-Barroso06, Bittner19}). Spatial binning was performed via the Voronoi tessellation technique \citep{Cappellari03} within the wavelength range of 480–580 nm, ensuring a target signal-to-noise ratio (S/N) of 40 per bin. Spaxels with S/N $\ge$ 40 remained unbinned, whereas those with S/N $\le$ 3 were excluded to eliminate noise contamination before binning. Furthermore, all spectra were corrected for galactic extinction along the line of sight to the target.

For the analysis of stellar kinematics, pPXF models the integrated observed spectrum of each binned unit as a non-negative linear combination of MILES spectral template libraries \citep{Vazdekis10}. The templates are convolved with the LOSVD in pixel space within the wavelength range 480 - 580 nm, as suggested in \citet{Bittner19, Bittner21}, to prevent contamination from the redder wavelength. The best-fit LOSVD parameters are determined through a least-squares minimisation procedure. Prominent spectral lines within the selected wavelength range are masked during the fitting process to minimise contamination. 
Based on the resulting best-fit parameters, spatially resolved maps of the stellar velocity ($V$) in the galaxy’s rest frame, velocity dispersion ($\sigma$), and higher-order Gauss-Hermite moments ($h_3$ and $h_4$) are generated.
\begin{figure*}
        \hspace{0.3cm}
        \centering
	\includegraphics[width=15 cm, height=13 cm]{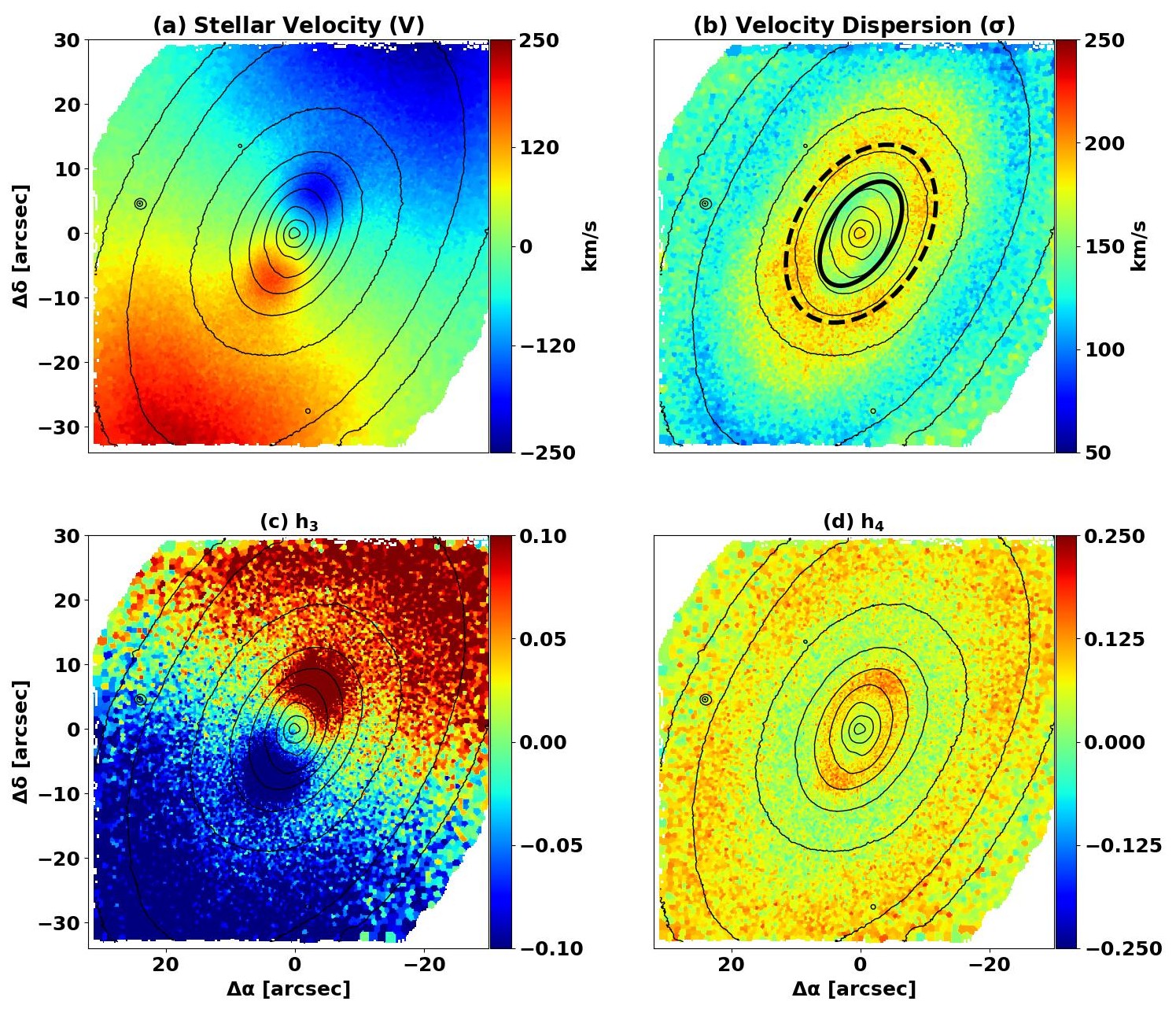}
    \caption{Stellar kinematic maps of NGC 1553 derived from the MUSE data cube using the GIST pipeline. The line-of-sight (LOS) stellar velocity ($V$) distribution is presented in the top-left panel, while the corresponding stellar velocity dispersion ($\sigma$) map is shown in the top-right panel. The nuclear disc is outlined by the solid black ellipse, while the dashed ellipse indicates the hot inner lens. The third ($h_{3}$) and fourth ($h_{4}$) Gauss-Hermite velocity moments are displayed in the bottom-left and bottom-right panels, respectively. The overlaid contours, extracted from the MUSE intensity maps, are spaced at uniform intervals of 0.2 mag. The orientation is such that the North is up, and the East is to the left.}
    \label{fig:Kinematics}
\end{figure*}
\begin{figure*}
    \centering
    \includegraphics[width=\textwidth]{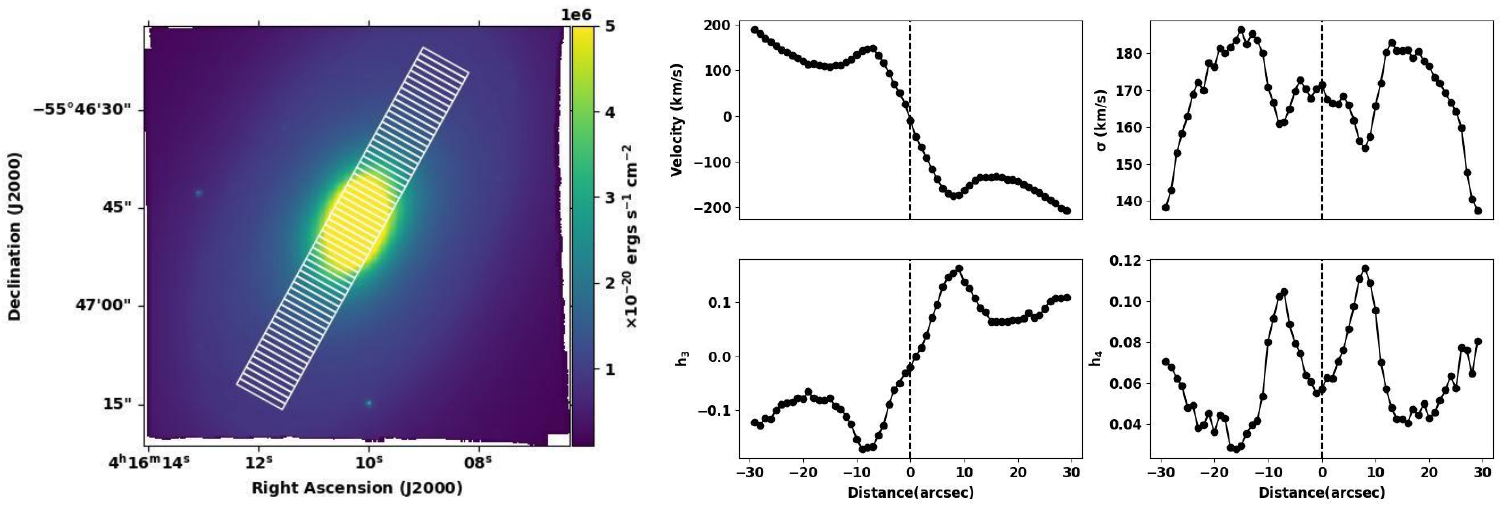}
    \caption{1D radial profiles from the corresponding 2D maps. \textbf{Left:} The intensity map of NGC 1553 is constructed using a MUSE datacube over the wavelength range of 480–580 nm. To extract the 1D profiles, we employed a rectangular pseudo-slit of width 8 arcsec and a step size of 1 arcsec along its length. \textbf{Right:} 1D radial profile of (a) $V$, (b) $\sigma$, (c) $h_{3}$ and (d) $h_{4}$ along the white rectangular pseudo-slit as shown on the left panel. The vertical dashed line in each profile signifies the position of the central slit. Typical median errors are $V_{err} \sim 6.1$ $km$ $s^{-1}$, $\sigma_{err} \sim 8.1$ $km$ $s^{-1}$, $h_{3,err} \sim 0.028$ and $h_{4,err} \sim 0.033$.}
    \label{fig:1Dprofile}
\end{figure*}


\section{Results}\label{sec:results}
\subsection{2D stellar kinematic maps}\label{sec:2D map}

In this section, we present the kinematic maps obtained from MUSE observations to investigate the signatures indicative of distinct structural components within the system. The higher-order moments, specifically the $h_{3}$ and $h_{4}$ coefficients, quantify the skewness and kurtosis of the LOSVD and serve as diagnostic tools for characterising the underlying stellar orbital structure. The LOSVD is extracted over the wavelength range of 480–580 nm. Figure~\ref{fig:Kinematics} presents the resulting stellar velocity ($V$), velocity dispersion ($\sigma$), as well as the $h_{3}$ and $h_{4}$ moment maps. The overlaid contours, derived from MUSE intensity maps, are spaced at uniform intervals of 0.2 mag.

The velocity ($V$) map (top left panel of Fig.~\ref{fig:Kinematics}) reveals a well-ordered rotation pattern characterised by redshifted velocities in the northwestern region of the galaxy and blueshifted velocities in the southeastern region. A distinct kinematic feature with an enhanced rotational component is evident within the central $\sim$ 8 arcsec. This component appears decoupled from the rest of the rotating disc, and suggests the existence of a nuclear ring, or disc. The radius of the ring corresponds to the radius of $\sim$ 8 arcsec ($\sim$ 0.62 kpc), where the dispersion is minimum, and the $V/\sigma$ is maximum.

The velocity dispersion ($\sigma$) map (top right panel of Fig.~\ref{fig:Kinematics}) exhibits a notable structure. A distinct, elongated structure with an increase in velocity dispersion is observed in the innermost region of the galaxy, aligned with the major axis of NGC 1553. Within the central 1.5 arcsec, $\sigma$ reaches $\sigma = 187 \pm 5$ km s$^{-1}$. This central peak in $\sigma$ is likely associated with a classical bulge/nuclear bar, as discussed by \citet{Erwin15}. Additionally, a decrease in velocity dispersion is detected at a radial distance of $\sim$ 8 arcsec, followed by a subsequent increase, reaching a maximum value of $\sigma = 183 \pm 7.7$ km s$^{-1}$ at a distance of 13 arcsec. Beyond this region, the velocity dispersion declines in the outer disc. We identified a faint ring-like structure with low $\sigma$ located approximately 30" from the center. This feature aligns with the outer ring detected by \citet{Rampazzo20} in their $H\alpha$ imaging, as shown in Figure C5 of their study.

The bottom-left panels of Fig.~\ref{fig:Kinematics} present the high-order Gauss-Hermite moment ${h_3}$, which characterises the skewness of the LOSVD. In the color map, red indicates positive skewness, corresponding to an excess of receding velocities relative to a purely Gaussian distribution. At the same time, blue denotes negative skewness, indicating an excess of approaching velocities. 
Correlations between $h_3$ and $V$ provide insights into the eccentricity of the underlying stellar orbits. An anti-correlation between $V$ and $h_3$ suggests the presence of near-circular orbits, which are characteristic of regularly rotating systems like stellar discs \citep{Bender94, Bureau05, Gadotti15, Bittner19, Gadotti20}. Such an anti-correlation is evident in the region of the inner disc, in particular at the central 8 arcsec region where the $V$ is high and $\sigma$ is low. In addition, this anti-correlation is observed across the galaxy, corresponding to the main disc of the galaxy. In contrast, a correlation between $V$ and $h_3$ indicates the presence of highly eccentric stellar orbits, like bars.

In the bottom-right panel of Fig.~\ref{fig:Kinematics}, we present the LOS kurtosis ($h_{4}$) map. The map exhibits spatially coherent structures, notably a region of elevated $h_{4}$ value. This enhancement in $h_{4}$ coincides with the observed anti-correlation between $V$ and $h_{3}$ in the central region. The presence of a high positive $h_{4}$ ring within the central 8-arcsec region suggests the superposition of multiple kinematic components with distinct LOSVDs \citep{Bender94}. In our analysis, this corresponds to regions of lower $\sigma$. We also observe another enhancement in $h_{4}$, appearing in a ring-like structure around 30". This feature corresponds spatially to the low $\sigma$ region detected at the same distance.

\begin{figure}
    \centering
    \includegraphics[width=8cm, height=5cm]{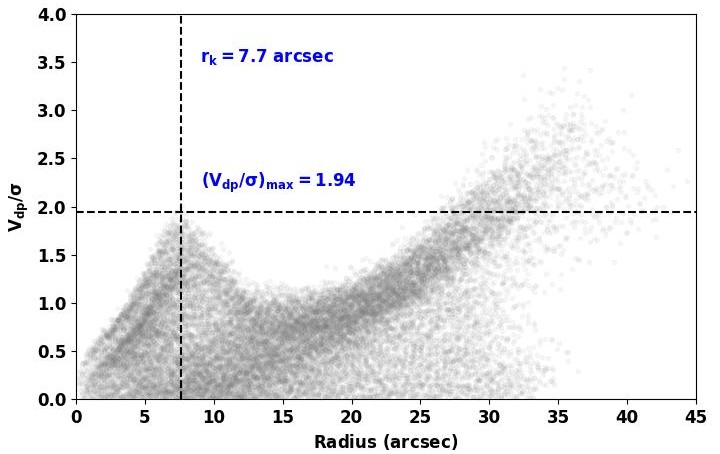}
    \caption{Deprojected radial profiles of $V/\sigma$ at each Voronoi bin, with velocity ($V$) corrected for inclination. The peak value $V_{dp}/\sigma$ and the radius at which this peak is located ($r_{k}$) are shown with the horizontal and vertical dashed lines, respectively, and the corresponding values are also given.}
    \label{fig:v/sigma}
\end{figure}

\subsection{1D radial profile}\label{sec:1D profile}
We derived one-dimensional (1D) radial profiles from the corresponding two-dimensional (2D) maps to facilitate a direct comparison with previous studies \citep{Kormendy84, Longo94, Fisher97, Chung04, Guerou16, Johnston18}. To extract these profiles, we employed a rectangular pseudo-slit, as illustrated in the intensity map (Fig.~\ref{fig:1Dprofile}). The slit has a width of 8 arcsec, approximately twice the minor-axis dimension of the nuclear disc, and extends sufficiently to encompass a significant portion of the galaxy's disc. This configuration enables the extraction of LOS velocity ($V$), velocity dispersion ($\sigma$), and $h_{3}$ and $h_{4}$ moments along the slit length. The slit was divided into multiple rectangular apertures of size 1 $\times$ 8 $arcsec^2$, with a step size of 1 arcsec along its length, as shown in Fig.\ref{fig:1Dprofile}). The vertical black dashed line marks the location of the galaxy's minor axis. All the features observed in 2D maps are perhaps better shown in the 1D line profiles. A clear anti-correlation is observed between $V$ (top left) and $h_3$ (bottom left) profiles throughout the galaxy. A small region of high positive and negative $h_3$ at the central 8 arcsec region corresponds to an anti-correlation with the $V$ profile. It appears to be decoupled from the rest of the galaxy, forming a dense and (quasi) axisymmetric central stellar disc \citep{Chung04}, this is also observed in most of the TIMER galaxies \citep{Gadotti20}. The 1D profile of $\sigma$ (top left) encodes detailed structural components of the central region. The peak in the central 1.5 arcsec is associated with the central classical bulge \citep{Erwin15}, followed by a drop in $\sigma$ around 8 arcsec associated with the nuclear disc co-located with the anti-correlation between $v$ and $h_3$. At a distance of $\sim13$ arcsec, the $\sigma$ increases again, reaching $183 \pm 7.7 ,\mathrm{km,s^{-1}}$, which corresponds to the hot inner lens. At the same radius, the LOS velocity decreases from its maximum value within the central $\sim8$ arcsec. Thus, the inner lens is dynamically hotter, as previously reported by \citet{Kormendy84}. Hereafter, we refer to the nuclear disc and the lens collectively as the nuclear disc–lens. The peak in the $h_4$ profile (bottom right) is co-located with regions of low $sigma$ and $V$ and $h_3$ anti-correlation.

To further analyse the kinematics, we computed the deprojected radial profiles of $V_{dp}/\sigma$ in each Voronoi bin (Fig.\ref{fig:Kinematics}). The maximum $V_{dp}/\sigma$ and the kinematic radius of the nuclear discs ($r_k$) are indicated in Fig.~\ref{fig:v/sigma} by the horizontal and vertical dashed lines, respectively. SK{The $r_k$ value is similar to the semimajor axis of the disc as determined using unsharp masking technique by \citep{Erwin15}.} The pronounced maximum of $V/\sigma$ indicates the presence of a nuclear disc, driven by the bar. Nuclear discs are kinematic components colder than the rest of their galaxy disc, as shown by \citet{Gadotti20}. The maximum value of $V/\sigma$ expected by simulations is around 2 \citep{Cole2014}.

\subsection{Stellar population properties}
\subsubsection{Non-parametric star formation history}\label{sec:sfh}

\begin{figure*}
    \centering
    \includegraphics[width=18cm, height=5.5cm]{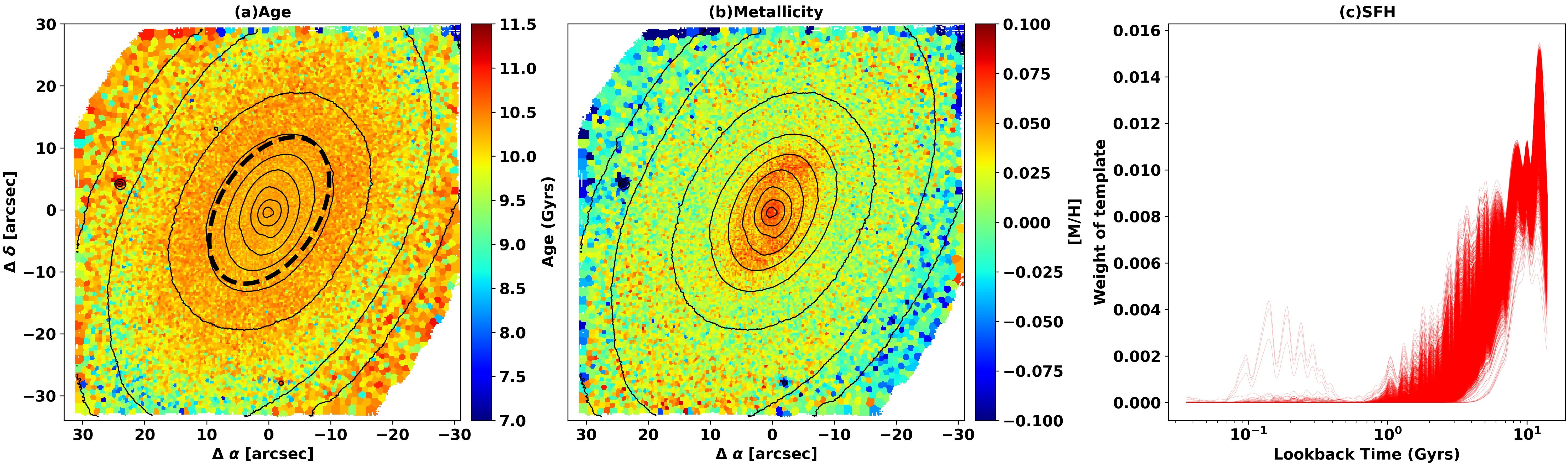}
    \caption{Mass-weighted stellar age map (left) and stellar metallicty map (middle) of NGC 1553 derived from the Voronoi-binned MUSE data cube. Overlaid black contours represent isophotes from the MUSE white-light image, spaced at intervals of 0.2 mag. The dashed black ellipse marks the central 13 arcsec region corresponding to the nuclear disc-lens. North is up and east is to the left. The star formation history (SFH) is shown on the right, where each profile represents the SFH of an individual bin.}
    \label{fig:age}
\end{figure*}

We derived the mass-weighted stellar populations and star-formation history (SFH) of NGC 1553 using SFH module of the GIST pipeline, which employs the pPXF algorithm to apply a regularized fit to the spectrum. It determines the underlying stellar population properties and recovers the associated non-parametric SFH through full spectral fitting \citep{Bittner19}. The analysis was performed on emission-line-subtracted spectra to ensure a robust continuum fit using template spectra. The fitting was conducted over the wavelength range 480–580 nm, incorporating an 8th-order multiplicative Legendre polynomial to account for discrepancies in the continuum shape between the observed and template spectra. The fit used a linear combination of MILES template library \citep{Vazdekis10}, which covers an age and metallicity range of 0.03 - 14 Gyr and [M/H] = -2.27 to +0.4, respectively. The models are based on the BaSTI isochrones \citep{Pietrinferni04, Pietrinferni06, Pietrinferni09, Pietrinferni13} and a revised Kroupa initial mass function \citep{Kroupa01}. The SFH and stellar population properties are estimated using the linear weights given to the template spectra. The resulting age (left) and metallicity (middle) are plotted in Fig. \ref{fig:age} along with the star formation history (right). The age distribution indicates that the galaxy hosts an old stellar population within the central region. The mean stellar age within the central $\sim$ 13 arcsec (within black dashed ellipse) is found to be 10.34 Gyr, suggesting that these structures formed in the early evolutionary stages of the galaxy. Additionally, in Fig. \ref{fig:age}, we observe an age gradient from the central to the outer regions of the galaxy. This suggests an inside-out formation of the disc. \citet{Johnston2021} found in a sample of S0 galaxies that the bulge and disc were comparable in age, and formed early from dissipational processes, while the lens was different, and formed later, in several episodes.

The metallicity map in Fig. \ref{fig:age} shows evidence of higher metallicity in the central region (within the region associated with the nuclear disc-lens) of the galaxy than in the outskirts. Similar metallicity trends are also observed in many earlier studies of S0s \citep{Ogando05, Bedregal11, Johnston2021}. \citet{Johnston2021} have also reported that lenses are generally metal-rich or have similar metallicity to discs, reflecting their in situ formation from the disc material. The old stellar population and higher metallicity in the lens of NGC 1553 suggest that most of the mass has been formed in the early stage of galaxy evolution, indicating that they may have formed in situ from the disc material.

\subsubsection{Single stellar population properties}\label{sec:lsf}

\begin{figure*}
    \centering
    \includegraphics[width=12cm, height=5.5cm]{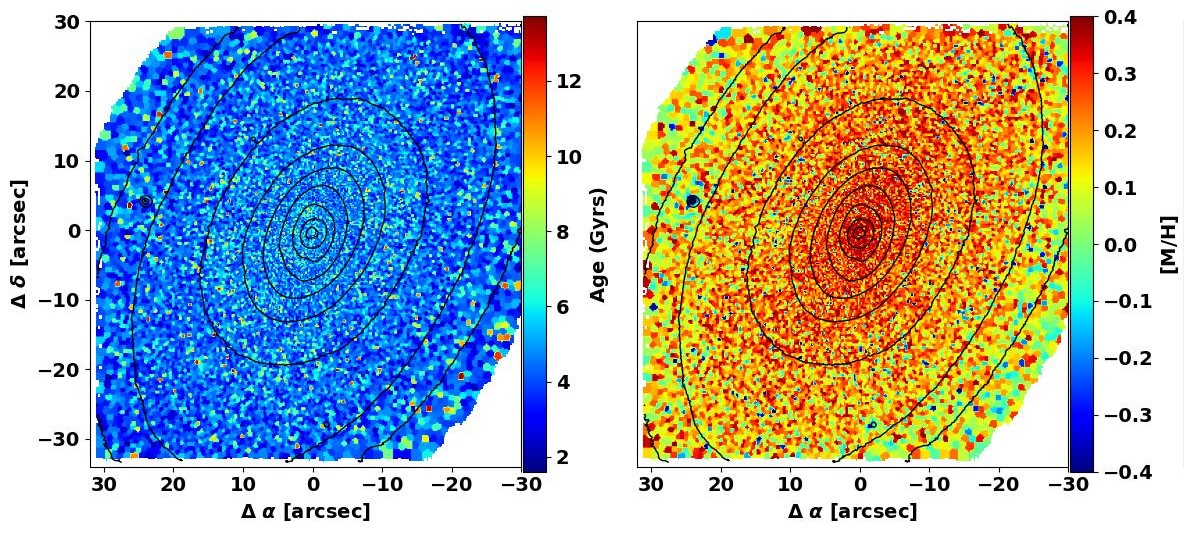}
    \caption{Maps of SSP-equivalent properties of NCG 1553 derived from absorption line strength measurements. Displayed are age (left) and [M/H] (right). Overlaid black contours represent isophotes from the MUSE white-light image, spaced at intervals of 0.2 mag. North is up and east is to the left.}
    \label{fig:ssp}
\end{figure*}

The stellar population properties derived using the line strength module of the GIST Pipeline include luminosity-weighted age, metallicity ([M/H]), and $\alpha$-enhancement ([$\alpha$/Fe]). We employed the LIS-8.4 system \citep{Vazdekis10} to measure absorption-line strength indices and convolved all spectra to a uniform spectral resolution of 8.4 Å, accounting for both instrumental and the local stellar velocity dispersion. We measured Lick indices for the hydrogen ($H\beta$), magnesium ($Mgb$), and iron ($Fe5270$ and $Fe5335$) \citep{Worthey94} to estimate single stellar population (SSP) properties. The luminosity-weighted stellar population is used to determine how long ago the recent star formation activity occurred. Thus, by analyzing the luminosity-weighted stellar populations, we can identify the regions of the galaxy where star formation was most recently truncated, offering insights into the processes that caused the gas to be stripped out and/or used up. The luminosity-weighted age and metallicities results are plotted as maps in Fig. \ref{fig:ssp}. The age maps (left panel of Fig. \ref{fig:ssp}) show a younger stellar population throughout the galaxy, which may indicate that the final episode of star formation occurred in the entire galaxy, resulting in a stellar population of age $\sim$ 4.5 Gyrs. This leads to the probability that the last stage of star formation in the galaxy is a result of some recent fly-by interaction or minor merger with satellite galaxies. S0s are generally observed to contain older stellar populations, though luminosity-weighted stellar population revealed a younger population of stars in several studies \citep{Kuntschner98, Poggianti01}. The metallicity map (right panel of Fig. \ref{fig:ssp}) reveals high metallicity within the central $\sim$13 arcseconds, beyond which the metallicity profile becomes largely flat with a small negative gradient, suggesting an accretion of low-metallicity ex-situ stars, likely through minor mergers \citep{Zibetti20}.

\section{Discussion \& Conclusion}\label{sec:discussion}
In this work, we present a detailed analysis of the kinematic and stellar population properties of NGC1~553. The unsharp-masked HST image of NGC~1553 confirms the presence of a nuclear bar and a faint spiral structure, as identified by \citet{Erwin15} (see their Fig.~A.1 for further details). These features lie within a nuclear disc that is embedded in a prominent inner lens; together we refer to this structure as the nuclear "disc-lens", which is itself surrounded by the main, large-scale disc.

Our detailed kinematic analysis using MUSE data reveals significant kinematic diversity. V/$\sigma$ map exhibits a central, rapidly rotating structure, suggesting a nuclear disc, encircling the nuclear bar. The $\sigma$ map displays prominent central peaks indicating a hot central region, and it reaches maxima at the inner lens position. $h_3$ shows a strong anti-correlation with the $V$, as expected and $h_4$ peaks notably at a radial distance of $\sim$8 arcsec, at the position of the nuclear disc.

The stellar population was analysed through both the regularised full spectral fitting and line strength measurement to obtain mass and luminosity-weighted properties, respectively. The mass-weighted age revealed that the galaxy's central region (specifically the lens) is old and metal-rich, whereas the disc comprises a mixed stellar population and is generally metal-poor. Additionally, we observe a small age gradient from the centre to the outskirts of the galaxy. Many studies \citep{Fraser-McKelvie18, Mendez-Abreu19, Johnston2021} found that the bulges in S0s are generally older and metal-rich than their disc, suggesting that star formation occurs throughout the discs but not in the bulge. The studies of spatially resolved stellar population properties using the CALIFA survey have revealed that the S0 galaxies are older than spirals, with ages ranging from 10.1-11.5 Gyr with a negative age gradient, where the centre is old and the outer disc is relatively younger \citep{Delgado15, Garcia17}. Such a negative age gradient reflects the inside-out growth of galaxies. While mass-weighted ages confirm the galaxy's disc mass built up over a relatively longer timescale than the central region, this is reminiscent of spiral galaxies, which typically host an older bulge and a star-forming disc. However, certain regions of the disc exhibit star formation timescales similar to those of the central region of the galaxy, suggesting that star formation was truncated relatively early in the galaxy, soon after the progenitor spiral was formed. The old and metal-rich stellar population in the central region (nuclear disc-lens) relative to the disc implies that the majority of the mass was created long ago via secular evolution.

Luminosity-weighted ages and metallicities derived from our SSP analysis point to a secondary, more recent episode of star formation that occurred after most of the galaxy’s stellar mass was built up. This rejuvenation can be an effect of a recent fly-by interaction or minor merger within the group environment, which could have funnelled fresh gas into the system and triggered star formation. The ionised gas distribution in NGC 1553 is notably diffuse and largely confined to the central regions of the galaxy. Within the inner ~8 arcseconds, it follows the spiral-like structure identified by \citet{Erwin15}. The line-of-sight (LOS) velocity of the ionised gas displays a rotation pattern broadly consistent with the stellar kinematics. Several studies have detected FUV emission in the central region of NGC 1553, along with clumpy $H\alpha$ emission extending over approximately 30 arcsec \citep{Rampazzo20, Rampazzo22}, indicating a low level of star formation in the galaxy, hence the presence of an underlying younger stellar population.

Observations indicate that lenses are more frequently found in the central regions of S0 galaxies \citep{Kormendy79, Laurikainen05, Laurikainen07, Laurikainen13, Gao18, Kruk18}. Various formation mechanisms have been proposed to explain the origin of lenses. \citet{Athanassoula83} suggested that lenses can emerge from a dynamically heated stellar population through gravitational instabilities, analogous to the formation of bars in disc galaxies.
 
Alternatively, \citet{Bournaud02, Bournaud05} proposed that lenses may form due to bar weakening induced by the inflow of cold gas. However, several studies have suggested that once a bar is established, it is inherently stable and difficult to dissolve \citep{Shen04}. Notably, S0 galaxies exhibit a lower bar fraction compared to spiral galaxies \citep{Laurikainen09, Barway11, Gao18}, suggesting a possible correlation between the presence of lenses and the absence of strong bars \citep{Kormendy79}. Numerical simulations have further demonstrated that lenses can also emerge through dissipationless mergers \citep{Eliche-Moral18}. These mergers have triggered bar formation or strengthened them, and accelerated their secular evolution. Approximately 58\% of merger remnant S0 galaxies exhibit identifiable inner structures, including embedded rings, pseudo-rings, inner spirals, discs, and nuclear bars, with nearly all containing a lens or an oval component—consistent with observations \citep{Laurikainen11, Laurikainen13}. These features may arise due to bar dilution and subsequent dynamical evolution \citep{Buta10, Laurikainen17}. \citet{Eliche-Moral18} argued that lenses and ovals in merger remnants are not solely a product of internal secular processes, but mergers may have fostered bars, highlighting the diversity in formation mechanisms.

To put NGC 1553 into context, we analysed the environment of galaxies from the \citet{Nair10} sample, and found that approximately 50\% of S0 galaxies with lenses are found in intermediate-density environments. This fraction is the same for barred spirals and also for barred S0 galaxies. The environment, therefore, does not play a key role in lens formation, nor in S0 formation.
Additionally, lenses are observed to be more prevalent in S0 galaxies than in spirals. S0 galaxies typically lack gas, so once a bar has weakened or dissipated in a lens, it remains in that state. In contrast,  Spiral galaxies, particularly late-type Spirals, can experience gas inflow, which may destabilize the disc and lead to the formation of a new bar, restarting the bar evolution cycle.

In a sample of 8 S0 galaxies, \citet{Johnston2021} found from stellar population and metallicity studies that bulges and discs formed at an early epoch from dissipational processes, while lenses formed later, at random times over the galaxy lifetime, possibly from evolved bars. They also find that field S0s have been more affected by minor mergers than galaxies in richer environments.

This study presents a comprehensive kinematic and stellar population analysis of the S0 galaxy NGC 1553, which hosts a lens and shell structures in a group environment. The galaxy exhibits a central rotating and dynamically hot nuclear disc-lens. Most of its stellar mass formed during the early stages of evolution. The nuclear disc-lens shows higher metallicity than the disc, indicating it likely formed in situ from disc material. Evidence of recent star formation suggests a recent interaction or minor merger event. Overall, the nuclear disc-lens in NGC 1553 appears to have formed early in the galaxy’s history, with its mass assembled over a shorter timescale than that of the disc. A detailed statistical and numerical study of lens-host S0 galaxies, particularly those in group environments, will provide further insights into the formation mechanisms of these kinematically distinct substructures in the S0's central regions.


\begin{acknowledgement}
We want to thank the anonymous referees for their useful and helpful feedback and comments on the previous version of this manuscript. SK thanks Dr. Chayan Mondal for the development of the 1d analysis methodology. This work uses observations collected at the European Southern Observatory under ESO program 0104.B-0404. It also uses the NASA/IPAC Extragalactic Database (NED), which is operated by the Jet Propulsion Laboratory, California Institute of Technology (Caltech), under contract with NASA. This research has made use of data obtained with the Dark Energy Camera (DECam), which was constructed by the Dark Energy Survey (DES) collaboration.

\end{acknowledgement}



\bibliographystyle{aa}
\bibliography{Reference}

\begin{thebibliography}{80}
\expandafter\ifx\csname natexlab\endcsname\relax\def\natexlab#1{#1}\fi

\bibitem[{{Athanassoula}(1983)}]{Athanassoula83}
{Athanassoula}, E., ed. 1983, IAU Symposium, Vol. 100, {Internal kinematics and dynamics of galaxies}

\bibitem[{{Athanassoula} {et~al.}(2016){Athanassoula}, {Rodionov}, {Peschken}, \& {Lambert}}]{Athanassoula16}
{Athanassoula}, E., {Rodionov}, S.~A., {Peschken}, N., \& {Lambert}, J.~C. 2016, \apj, 821, 90

\bibitem[{{Bacon} {et~al.}(2006){Bacon}, {Bauer}, {Boehm}, {Boudon}, {Brau-Nogu{\'e}}, {Caillier}, {Capoani}, {Carollo}, {Champavert}, {Contini}, {Daguis{\'e}}, {Dall{\'e}}, {Delabre}, {Devriendt}, {Dreizler}, {Dubois}, {Dupieux}, {Dupin}, {Emsellem}, {Ferruit}, {Franx}, {Gallou}, {Gerssen}, {Guiderdoni}, {Hahn}, {Hofmann}, {Jarno}, {Kelz}, {Koehler}, {Kollatschny}, {Kosmalski}, {Laurent}, {Lilly}, {Lizon}, {Loupias}, {Lynn}, {Manescau}, {McDermid}, {Monstein}, {Nicklas}, {Par{\`e}s}, {Pasquini}, {P{\'e}contal-Rousset}, {P{\'e}contal}, {Pello}, {Petit}, {Picat}, {Popow}, {Quirrenbach}, {Reiss}, {Renault}, {Roth}, {Schaye}, {Soucail}, {Steinmetz}, {Stroebele}, {Stuik}, {Weilbacher}, {Wozniak}, \& {de Zeeuw}}]{Bacon06}
{Bacon}, R., {Bauer}, S., {Boehm}, P., {et~al.} 2006, in Society of Photo-Optical Instrumentation Engineers (SPIE) Conference Series, Vol. 6269, Ground-based and Airborne Instrumentation for Astronomy, ed. I.~S. {McLean} \& M.~{Iye}, 62690J

\bibitem[{{Barway} {et~al.}(2011){Barway}, {Wadadekar}, \& {Kembhavi}}]{Barway11}
{Barway}, S., {Wadadekar}, Y., \& {Kembhavi}, A.~K. 2011, \mnras, 410, L18

\bibitem[{Bedregal {et~al.}(2011)Bedregal, Cardiel, Aragón-Salamanca, \& Merrifield}]{Bedregal11}
Bedregal, A.~G., Cardiel, N., Aragón-Salamanca, A., \& Merrifield, M.~R. 2011, Monthly Notices of the Royal Astronomical Society, 415, 2063

\bibitem[{{Bekki} \& {Couch}(2011)}]{Bekki11}
{Bekki}, K. \& {Couch}, W.~J. 2011, \mnras, 415, 1783

\bibitem[{{Bender} {et~al.}(1994){Bender}, {Saglia}, \& {Gerhard}}]{Bender94}
{Bender}, R., {Saglia}, R.~P., \& {Gerhard}, O.~E. 1994, \mnras, 269, 785

\bibitem[{{Bittner} {et~al.}(2021){Bittner}, {de Lorenzo-C{\'a}ceres}, {Gadotti}, {S{\'a}nchez-Bl{\'a}zquez}, {Neumann}, {Coelho}, {Falc{\'o}n-Barroso}, {Fragkoudi}, {Kim}, {Mart{\'\i}n-Navarro}, {M{\'e}ndez-Abreu}, {P{\'e}rez}, {Querejeta}, \& {van de Ven}}]{Bittner21}
{Bittner}, A., {de Lorenzo-C{\'a}ceres}, A., {Gadotti}, D.~A., {et~al.} 2021, \aap, 646, A42

\bibitem[{{Bittner} {et~al.}(2019){Bittner}, {Falc{\'o}n-Barroso}, {Nedelchev}, {Dorta}, {Gadotti}, {Sarzi}, {Molaeinezhad}, {Iodice}, {Rosado-Belza}, {de Lorenzo-C{\'a}ceres}, {Fragkoudi}, {Gal{\'a}n-de Anta}, {Husemann}, {M{\'e}ndez-Abreu}, {Neumann}, {Pinna}, {Querejeta}, {S{\'a}nchez-Bl{\'a}zquez}, \& {Seidel}}]{Bittner19}
{Bittner}, A., {Falc{\'o}n-Barroso}, J., {Nedelchev}, B., {et~al.} 2019, \aap, 628, A117

\bibitem[{{Blakeslee} {et~al.}(2001){Blakeslee}, {Lucey}, {Barris}, {Hudson}, \& {Tonry}}]{Blakeslee01}
{Blakeslee}, J.~P., {Lucey}, J.~R., {Barris}, B.~J., {Hudson}, M.~J., \& {Tonry}, J.~L. 2001, \mnras, 327, 1004

\bibitem[{{Blanton} {et~al.}(2001){Blanton}, {Sarazin}, \& {Irwin}}]{Blanton01}
{Blanton}, E.~L., {Sarazin}, C.~L., \& {Irwin}, J.~A. 2001, \apj, 552, 106

\bibitem[{{Bournaud} \& {Combes}(2002)}]{Bournaud02}
{Bournaud}, F. \& {Combes}, F. 2002, \aap, 392, 83

\bibitem[{{Bournaud} {et~al.}(2005){Bournaud}, {Combes}, \& {Semelin}}]{Bournaud05}
{Bournaud}, F., {Combes}, F., \& {Semelin}, B. 2005, \mnras, 364, L18

\bibitem[{{Bridges} \& {Hanes}(1990)}]{Bridges90}
{Bridges}, T.~J. \& {Hanes}, D.~A. 1990, \aj, 99, 1100

\bibitem[{{Bureau} \& {Athanassoula}(2005)}]{Bureau05}
{Bureau}, M. \& {Athanassoula}, E. 2005, \apj, 626, 159

\bibitem[{{Buta} {et~al.}(2010){Buta}, {Laurikainen}, {Salo}, \& {Knapen}}]{Buta10}
{Buta}, R., {Laurikainen}, E., {Salo}, H., \& {Knapen}, J.~H. 2010, \apj, 721, 259

\bibitem[{{Buta} {et~al.}(2015){Buta}, {Sheth}, {Athanassoula}, {Bosma}, {Knapen}, {Laurikainen}, {Salo}, {Elmegreen}, {Ho}, {Zaritsky}, {Courtois}, {Hinz}, {Mu{\~n}oz-Mateos}, {Kim}, {Regan}, {Gadotti}, {Gil de Paz}, {Laine}, {Men{\'e}ndez-Delmestre}, {Comer{\'o}n}, {Erroz Ferrer}, {Seibert}, {Mizusawa}, {Holwerda}, \& {Madore}}]{Buta15}
{Buta}, R.~J., {Sheth}, K., {Athanassoula}, E., {et~al.} 2015, \apjs, 217, 32

\bibitem[{{Cappellari}(2017)}]{Cappellari17}
{Cappellari}, M. 2017, \mnras, 466, 798

\bibitem[{{Cappellari} \& {Copin}(2003)}]{Cappellari03}
{Cappellari}, M. \& {Copin}, Y. 2003, \mnras, 342, 345

\bibitem[{{Cappellari} \& {Emsellem}(2004)}]{Cappellari04}
{Cappellari}, M. \& {Emsellem}, E. 2004, \pasp, 116, 138

\bibitem[{{Chapon} {et~al.}(2013){Chapon}, {Mayer}, \& {Teyssier}}]{Chapon13}
{Chapon}, D., {Mayer}, L., \& {Teyssier}, R. 2013, \mnras, 429, 3114

\bibitem[{{Chung} \& {Bureau}(2004)}]{Chung04}
{Chung}, A. \& {Bureau}, M. 2004, \aj, 127, 3192

\bibitem[{{Cole} {et~al.}(2014){Cole}, {Debattista}, {Erwin}, {Earp}, \& {Ro{\v{s}}kar}}]{Cole2014}
{Cole}, D.~R., {Debattista}, V.~P., {Erwin}, P., {Earp}, S. W.~F., \& {Ro{\v{s}}kar}, R. 2014, \mnras, 445, 3352

\bibitem[{{Combes}(1996)}]{Combes96}
{Combes}, F. 1996, in Astronomical Society of the Pacific Conference Series, Vol.~91, IAU Colloq. 157: Barred Galaxies, ed. R.~{Buta}, D.~A. {Crocker}, \& B.~G. {Elmegreen}, 286

\bibitem[{{de Vaucouleurs} {et~al.}(1991){de Vaucouleurs}, {de Vaucouleurs}, {Corwin}, {Buta}, {Paturel}, \& {Fouque}}]{deVaucouleurs91}
{de Vaucouleurs}, G., {de Vaucouleurs}, A., {Corwin}, Herold~G., J., {et~al.} 1991, {Third Reference Catalogue of Bright Galaxies}

\bibitem[{{Dey} {et~al.}(2019){Dey}, {Schlegel}, {Lang}, {Blum}, {Burleigh}, {Fan}, {Findlay}, {Finkbeiner}, {Herrera}, {Juneau}, {Landriau}, {Levi}, {McGreer}, {Meisner}, {Myers}, {Moustakas}, {Nugent}, {Patej}, {Schlafly}, {Walker}, {Valdes}, {Weaver}, {Y{\`e}che}, {Zou}, {Zhou}, {Abareshi}, {Abbott}, {Abolfathi}, {Aguilera}, {Alam}, {Allen}, {Alvarez}, {Annis}, {Ansarinejad}, {Aubert}, {Beechert}, {Bell}, {BenZvi}, {Beutler}, {Bielby}, {Bolton}, {Brice{\~n}o}, {Buckley-Geer}, {Butler}, {Calamida}, {Carlberg}, {Carter}, {Casas}, {Castander}, {Choi}, {Comparat}, {Cukanovaite}, {Delubac}, {DeVries}, {Dey}, {Dhungana}, {Dickinson}, {Ding}, {Donaldson}, {Duan}, {Duckworth}, {Eftekharzadeh}, {Eisenstein}, {Etourneau}, {Fagrelius}, {Farihi}, {Fitzpatrick}, {Font-Ribera}, {Fulmer}, {G{\"a}nsicke}, {Gaztanaga}, {George}, {Gerdes}, {Gontcho}, {Gorgoni}, {Green}, {Guy}, {Harmer}, {Hernandez}, {Honscheid}, {Huang}, {James}, {Jannuzi}, {Jiang}, {Joyce}, {Karcher}, {Karkar}, {Kehoe}, {Kneib}, {Kueter-Young}, {Lan},
  {Lauer}, {Le Guillou}, {Le Van Suu}, {Lee}, {Lesser}, {Perreault Levasseur}, {Li}, {Mann}, {Marshall}, {Mart{\'\i}nez-V{\'a}zquez}, {Martini}, {du Mas des Bourboux}, {McManus}, {Meier}, {M{\'e}nard}, {Metcalfe}, {Mu{\~n}oz-Guti{\'e}rrez}, {Najita}, {Napier}, {Narayan}, {Newman}, {Nie}, {Nord}, {Norman}, {Olsen}, {Paat}, {Palanque-Delabrouille}, {Peng}, {Poppett}, {Poremba}, {Prakash}, {Rabinowitz}, {Raichoor}, {Rezaie}, {Robertson}, {Roe}, {Ross}, {Ross}, {Rudnick}, {Safonova}, {Saha}, {S{\'a}nchez}, {Savary}, {Schweiker}, {Scott}, {Seo}, {Shan}, {Silva}, {Slepian}, {Soto}, {Sprayberry}, {Staten}, {Stillman}, {Stupak}, {Summers}, {Sien Tie}, {Tirado}, {Vargas-Maga{\~n}a}, {Vivas}, {Wechsler}, {Williams}, {Yang}, {Yang}, {Yapici}, {Zaritsky}, {Zenteno}, {Zhang}, {Zhang}, {Zhou}, \& {Zhou}}]{Dey19}
{Dey}, A., {Schlegel}, D.~J., {Lang}, D., {et~al.} 2019, \aj, 157, 168

\bibitem[{{Eliche-Moral} {et~al.}(2018){Eliche-Moral}, {Rodr{\'\i}guez-P{\'e}rez}, {Borlaff}, {Querejeta}, \& {Tapia}}]{Eliche-Moral18}
{Eliche-Moral}, M.~C., {Rodr{\'\i}guez-P{\'e}rez}, C., {Borlaff}, A., {Querejeta}, M., \& {Tapia}, T. 2018, \aap, 617, A113

\bibitem[{{Erwin} {et~al.}(2015){Erwin}, {Saglia}, {Fabricius}, {Thomas}, {Nowak}, {Rusli}, {Bender}, {Vega Beltr{\'a}n}, \& {Beckman}}]{Erwin15}
{Erwin}, P., {Saglia}, R.~P., {Fabricius}, M., {et~al.} 2015, \mnras, 446, 4039

\bibitem[{{Falc{\'o}n-Barroso} {et~al.}(2006{\natexlab{a}}){Falc{\'o}n-Barroso}, {Bacon}, {Bureau}, {Cappellari}, {Davies}, {de Zeeuw}, {Emsellem}, {Fathi}, {Krajnovi{\'c}}, {Kuntschner}, {McDermid}, {Peletier}, \& {Sarzi}}]{Falcon06}
{Falc{\'o}n-Barroso}, J., {Bacon}, R., {Bureau}, M., {et~al.} 2006{\natexlab{a}}, \mnras, 369, 529

\bibitem[{{Falc{\'o}n-Barroso} {et~al.}(2006{\natexlab{b}}){Falc{\'o}n-Barroso}, {Bacon}, {Bureau}, {Cappellari}, {Davies}, {de Zeeuw}, {Emsellem}, {Fathi}, {Krajnovi{\'c}}, {Kuntschner}, {McDermid}, {Peletier}, \& {Sarzi}}]{Falc'on-Barroso06}
{Falc{\'o}n-Barroso}, J., {Bacon}, R., {Bureau}, M., {et~al.} 2006{\natexlab{b}}, \mnras, 369, 529

\bibitem[{{Falc{\'o}n-Barroso} {et~al.}(2004){Falc{\'o}n-Barroso}, {Peletier}, {Emsellem}, {Kuntschner}, {Fathi}, {Bureau}, {Bacon}, {Cappellari}, {Copin}, {Davies}, \& {de Zeeuw}}]{Falcon04}
{Falc{\'o}n-Barroso}, J., {Peletier}, R.~F., {Emsellem}, E., {et~al.} 2004, \mnras, 350, 35

\bibitem[{{Fisher}(1997)}]{Fisher97}
{Fisher}, D. 1997, \aj, 113, 950

\bibitem[{{Fraser-McKelvie} {et~al.}(2018){Fraser-McKelvie}, {Arag{\'o}n-Salamanca}, {Merrifield}, {Tabor}, {Bernardi}, {Drory}, {Parikh}, \& {Argudo-Fern{\'a}ndez}}]{Fraser-McKelvie18}
{Fraser-McKelvie}, A., {Arag{\'o}n-Salamanca}, A., {Merrifield}, M., {et~al.} 2018, \mnras, 481, 5580

\bibitem[{{Gadotti} {et~al.}(2020){Gadotti}, {Bittner}, {Falc{\'o}n-Barroso}, {M{\'e}ndez-Abreu}, {Kim}, {Fragkoudi}, {de Lorenzo-C{\'a}ceres}, {Leaman}, {Neumann}, {Querejeta}, {S{\'a}nchez-Bl{\'a}zquez}, {Martig}, {Mart{\'\i}n-Navarro}, {P{\'e}rez}, {Seidel}, \& {van de Ven}}]{Gadotti20}
{Gadotti}, D.~A., {Bittner}, A., {Falc{\'o}n-Barroso}, J., {et~al.} 2020, \aap, 643, A14

\bibitem[{{Gadotti} {et~al.}(2015){Gadotti}, {Seidel}, {S{\'a}nchez-Bl{\'a}zquez}, {Falc{\'o}n-Barroso}, {Husemann}, {Coelho}, \& {P{\'e}rez}}]{Gadotti15}
{Gadotti}, D.~A., {Seidel}, M.~K., {S{\'a}nchez-Bl{\'a}zquez}, P., {et~al.} 2015, \aap, 584, A90

\bibitem[{{Gao} {et~al.}(2018){Gao}, {Ho}, {Barth}, \& {Li}}]{Gao18}
{Gao}, H., {Ho}, L.~C., {Barth}, A.~J., \& {Li}, Z.-Y. 2018, \apj, 862, 100

\bibitem[{{Garcia}(1993)}]{Garcia93}
{Garcia}, A.~M. 1993, \aaps, 100, 47

\bibitem[{{Garc{\'\i}a-Benito} {et~al.}(2017){Garc{\'\i}a-Benito}, {Gonz{\'a}lez Delgado}, {P{\'e}rez}, {Cid Fernandes}, {Cortijo-Ferrero}, {L{\'o}pez Fern{\'a}ndez}, {de Amorim}, {Lacerda}, {Vale Asari}, \& {S{\'a}nchez}}]{Garcia17}
{Garc{\'\i}a-Benito}, R., {Gonz{\'a}lez Delgado}, R.~M., {P{\'e}rez}, E., {et~al.} 2017, \aap, 608, A27

\bibitem[{{Giri} {et~al.}(2023){Giri}, {Barway}, \& {Raychaudhury}}]{Gourab23}
{Giri}, G., {Barway}, S., \& {Raychaudhury}, S. 2023, \mnras, 520, 5870

\bibitem[{{Gonz{\'a}lez Delgado} {et~al.}(2015){Gonz{\'a}lez Delgado}, {Garc{\'\i}a-Benito}, {P{\'e}rez}, {Cid Fernandes}, {de Amorim}, {Cortijo-Ferrero}, {Lacerda}, {L{\'o}pez Fern{\'a}ndez}, {Vale-Asari}, {S{\'a}nchez}, {Moll{\'a}}, {Ruiz-Lara}, {S{\'a}nchez-Bl{\'a}zquez}, {Walcher}, {Alves}, {Aguerri}, {Bekerait{\'e}}, {Bland-Hawthorn}, {Galbany}, {Gallazzi}, {Husemann}, {Iglesias-P{\'a}ramo}, {Kalinova}, {L{\'o}pez-S{\'a}nchez}, {Marino}, {M{\'a}rquez}, {Masegosa}, {Mast}, {M{\'e}ndez-Abreu}, {Mendoza}, {del Olmo}, {P{\'e}rez}, {Quirrenbach}, \& {Zibetti}}]{Delgado15}
{Gonz{\'a}lez Delgado}, R.~M., {Garc{\'\i}a-Benito}, R., {P{\'e}rez}, E., {et~al.} 2015, \aap, 581, A103

\bibitem[{{Gu{\'e}rou} {et~al.}(2016){Gu{\'e}rou}, {Emsellem}, {Krajnovi{\'c}}, {McDermid}, {Contini}, \& {Weilbacher}}]{Guerou16}
{Gu{\'e}rou}, A., {Emsellem}, E., {Krajnovi{\'c}}, D., {et~al.} 2016, \aap, 591, A143

\bibitem[{{Johnston} {et~al.}(2021){Johnston}, {Arag{\'o}n-Salamanca}, {Fraser-McKelvie}, {Merrifield}, {H{\"a}u{\ss}ler}, {Coccato}, {Jaff{\'e}}, {Cortesi}, {Chies-Santos}, {Rodr{\'\i}guez Del Pino}, \& {Sheen}}]{Johnston2021}
{Johnston}, E.~J., {Arag{\'o}n-Salamanca}, A., {Fraser-McKelvie}, A., {et~al.} 2021, \mnras, 500, 4193

\bibitem[{{Johnston} {et~al.}(2018){Johnston}, {Hau}, {Coccato}, \& {Herrera}}]{Johnston18}
{Johnston}, E.~J., {Hau}, G. K.~T., {Coccato}, L., \& {Herrera}, C. 2018, \mnras, 480, 3215

\bibitem[{{Keshri} {et~al.}(2025){Keshri}, {Barway}, {Das}, {Yadav}, \& {Combes}}]{Saili25}
{Keshri}, S., {Barway}, S., {Das}, M., {Yadav}, J., \& {Combes}, F. 2025, arXiv e-prints, arXiv:2501.08544

\bibitem[{{Kormendy}(1979)}]{Kormendy79}
{Kormendy}, J. 1979, \apj, 227, 714

\bibitem[{{Kormendy}(1984)}]{Kormendy84}
{Kormendy}, J. 1984, \apj, 286, 116

\bibitem[{{Kormendy} \& {Kennicutt}(2004)}]{Kormendy04}
{Kormendy}, J. \& {Kennicutt}, Robert~C., J. 2004, \araa, 42, 603

\bibitem[{{Kroupa}(2001)}]{Kroupa01}
{Kroupa}, P. 2001, \mnras, 322, 231

\bibitem[{{Kruk} {et~al.}(2018){Kruk}, {Lintott}, {Bamford}, {Masters}, {Simmons}, {H{\"a}u{\ss}ler}, {Cardamone}, {Hart}, {Kelvin}, {Schawinski}, {Smethurst}, \& {Vika}}]{Kruk18}
{Kruk}, S.~J., {Lintott}, C.~J., {Bamford}, S.~P., {et~al.} 2018, \mnras, 473, 4731

\bibitem[{{Kuntschner} \& {Davies}(1998)}]{Kuntschner98}
{Kuntschner}, H. \& {Davies}, R.~L. 1998, \mnras, 295, L29

\bibitem[{{Laurikainen} \& {Salo}(2017)}]{Laurikainen17}
{Laurikainen}, E. \& {Salo}, H. 2017, \aap, 598, A10

\bibitem[{{Laurikainen} {et~al.}(2013){Laurikainen}, {Salo}, {Athanassoula}, {Bosma}, {Buta}, \& {Janz}}]{Laurikainen13}
{Laurikainen}, E., {Salo}, H., {Athanassoula}, E., {et~al.} 2013, \mnras, 430, 3489

\bibitem[{{Laurikainen} {et~al.}(2005){Laurikainen}, {Salo}, \& {Buta}}]{Laurikainen05}
{Laurikainen}, E., {Salo}, H., \& {Buta}, R. 2005, \mnras, 362, 1319

\bibitem[{{Laurikainen} {et~al.}(2006){Laurikainen}, {Salo}, {Buta}, {Knapen}, {Speltincx}, \& {Block}}]{Laurikainen06}
{Laurikainen}, E., {Salo}, H., {Buta}, R., {et~al.} 2006, \aj, 132, 2634

\bibitem[{{Laurikainen} {et~al.}(2007){Laurikainen}, {Salo}, {Buta}, \& {Knapen}}]{Laurikainen07}
{Laurikainen}, E., {Salo}, H., {Buta}, R., \& {Knapen}, J.~H. 2007, \mnras, 381, 401

\bibitem[{{Laurikainen} {et~al.}(2009){Laurikainen}, {Salo}, {Buta}, \& {Knapen}}]{Laurikainen09}
{Laurikainen}, E., {Salo}, H., {Buta}, R., \& {Knapen}, J.~H. 2009, \apjl, 692, L34

\bibitem[{{Laurikainen} {et~al.}(2011){Laurikainen}, {Salo}, {Buta}, \& {Knapen}}]{Laurikainen11}
{Laurikainen}, E., {Salo}, H., {Buta}, R., \& {Knapen}, J.~H. 2011, \mnras, 418, 1452

\bibitem[{{Longo} {et~al.}(1994){Longo}, {Zaggia}, {Busarello}, \& {Richter}}]{Longo94}
{Longo}, G., {Zaggia}, S.~R., {Busarello}, G., \& {Richter}, G. 1994, \aaps, 105, 433

\bibitem[{{Malin}(1978)}]{Malin78}
{Malin}, D.~F. 1978, \nat, 276, 591

\bibitem[{{Malin} \& {Carter}(1983)}]{Malin83}
{Malin}, D.~F. \& {Carter}, D. 1983, \apj, 274, 534

\bibitem[{{Mayer} {et~al.}(2010){Mayer}, {Kazantzidis}, {Escala}, \& {Callegari}}]{Mayer10}
{Mayer}, L., {Kazantzidis}, S., {Escala}, A., \& {Callegari}, S. 2010, \nat, 466, 1082

\bibitem[{{M{\'e}ndez-Abreu} {et~al.}(2019){M{\'e}ndez-Abreu}, {S{\'a}nchez}, \& {de Lorenzo-C{\'a}ceres}}]{Mendez-Abreu19}
{M{\'e}ndez-Abreu}, J., {S{\'a}nchez}, S.~F., \& {de Lorenzo-C{\'a}ceres}, A. 2019, \mnras, 488, L80

\bibitem[{{Nair} \& {Abraham}(2010)}]{Nair10}
{Nair}, P.~B. \& {Abraham}, R.~G. 2010, \apjs, 186, 427

\bibitem[{{Ogando} {et~al.}(2005){Ogando}, {Maia}, {Chiappini}, {Pellegrini}, {Schiavon}, \& {da Costa}}]{Ogando05}
{Ogando}, R. L.~C., {Maia}, M. A.~G., {Chiappini}, C., {et~al.} 2005, \apjl, 632, L61

\bibitem[{{Pietrinferni} {et~al.}(2004){Pietrinferni}, {Cassisi}, {Salaris}, \& {Castelli}}]{Pietrinferni04}
{Pietrinferni}, A., {Cassisi}, S., {Salaris}, M., \& {Castelli}, F. 2004, \apj, 612, 168

\bibitem[{{Pietrinferni} {et~al.}(2006){Pietrinferni}, {Cassisi}, {Salaris}, \& {Castelli}}]{Pietrinferni06}
{Pietrinferni}, A., {Cassisi}, S., {Salaris}, M., \& {Castelli}, F. 2006, \apj, 642, 797

\bibitem[{{Pietrinferni} {et~al.}(2013){Pietrinferni}, {Cassisi}, {Salaris}, \& {Hidalgo}}]{Pietrinferni13}
{Pietrinferni}, A., {Cassisi}, S., {Salaris}, M., \& {Hidalgo}, S. 2013, \aap, 558, A46

\bibitem[{{Pietrinferni} {et~al.}(2009){Pietrinferni}, {Cassisi}, {Salaris}, {Percival}, \& {Ferguson}}]{Pietrinferni09}
{Pietrinferni}, A., {Cassisi}, S., {Salaris}, M., {Percival}, S., \& {Ferguson}, J.~W. 2009, \apj, 697, 275

\bibitem[{{Poggianti} {et~al.}(2001){Poggianti}, {Bridges}, {Carter}, {Mobasher}, {Doi}, {Iye}, {Kashikawa}, {Komiyama}, {Okamura}, {Sekiguchi}, {Shimasaku}, {Yagi}, \& {Yasuda}}]{Poggianti01}
{Poggianti}, B.~M., {Bridges}, T.~J., {Carter}, D., {et~al.} 2001, \apj, 563, 118

\bibitem[{{Rampazzo} {et~al.}(2020){Rampazzo}, {Ciroi}, {Mazzei}, {Di Mille}, {Congiu}, {Cattapan}, {Bianchi}, {Iodice}, {Marino}, {Plana}, {Postma}, \& {Spavone}}]{Rampazzo20}
{Rampazzo}, R., {Ciroi}, S., {Mazzei}, P., {et~al.} 2020, \aap, 643, A176

\bibitem[{{Rampazzo} {et~al.}(2022){Rampazzo}, {Mazzei}, {Marino}, {Bianchi}, {Postma}, {Ragusa}, {Spavone}, {Iodice}, {Ciroi}, \& {Held}}]{Rampazzo22}
{Rampazzo}, R., {Mazzei}, P., {Marino}, A., {et~al.} 2022, \aap, 664, A192

\bibitem[{{Ricci} {et~al.}(2023){Ricci}, {Steiner}, {Menezes}, {Slodkowski Clerici}, \& {da Silva}}]{Ricci23}
{Ricci}, T.~V., {Steiner}, J.~E., {Menezes}, R.~B., {Slodkowski Clerici}, K., \& {da Silva}, M.~D. 2023, \mnras, 522, 2207

\bibitem[{{Sandage} \& {Brucato}(1979)}]{Sandage79}
{Sandage}, A. \& {Brucato}, R. 1979, \aj, 84, 472

\bibitem[{{Sarzi} {et~al.}(2006){Sarzi}, {Falc{\'o}n-Barroso}, {Davies}, {Bacon}, {Bureau}, {Cappellari}, {de Zeeuw}, {Emsellem}, {Fathi}, {Krajnovi{\'c}}, {Kuntschner}, {McDermid}, \& {Peletier}}]{Sarzi06}
{Sarzi}, M., {Falc{\'o}n-Barroso}, J., {Davies}, R.~L., {et~al.} 2006, \mnras, 366, 1151

\bibitem[{{Shen} \& {Sellwood}(2004)}]{Shen04}
{Shen}, J. \& {Sellwood}, J.~A. 2004, \apj, 604, 614

\bibitem[{{Trinchieri} {et~al.}(1997){Trinchieri}, {Noris}, \& {di Serego Alighieri}}]{Trinch97}
{Trinchieri}, G., {Noris}, L., \& {di Serego Alighieri}, S. 1997, \aap, 326, 565

\bibitem[{{Vazdekis} {et~al.}(2010){Vazdekis}, {S{\'a}nchez-Bl{\'a}zquez}, {Falc{\'o}n-Barroso}, {Cenarro}, {Beasley}, {Cardiel}, {Gorgas}, \& {Peletier}}]{Vazdekis10}
{Vazdekis}, A., {S{\'a}nchez-Bl{\'a}zquez}, P., {Falc{\'o}n-Barroso}, J., {et~al.} 2010, \mnras, 404, 1639

\bibitem[{{Wilman} {et~al.}(2009){Wilman}, {Oemler}, {Mulchaey}, {McGee}, {Balogh}, \& {Bower}}]{Wilman09}
{Wilman}, D.~J., {Oemler}, Jr., A., {Mulchaey}, J.~S., {et~al.} 2009, \apj, 692, 298

\bibitem[{{Worthey} {et~al.}(1994){Worthey}, {Faber}, {Gonzalez}, \& {Burstein}}]{Worthey94}
{Worthey}, G., {Faber}, S.~M., {Gonzalez}, J.~J., \& {Burstein}, D. 1994, \apjs, 94, 687

\bibitem[{{Zibetti} {et~al.}(2020){Zibetti}, {Gallazzi}, {Hirschmann}, {Consolandi}, {Falc{\'o}n-Barroso}, {van de Ven}, \& {Lyubenova}}]{Zibetti20}
{Zibetti}, S., {Gallazzi}, A.~R., {Hirschmann}, M., {et~al.} 2020, \mnras, 491, 3562

\end{thebibliography}

\end{document}